\documentclass[12pt]{article}
\usepackage{graphicx}
\usepackage{amsbsy}

\title{Unfolding of eigenvalue surfaces near a diabolic point due to a complex perturbation}
\author{O.~N.~Kirillov, A.~A.~Mailybaev, and A.~P.~Seyranian}

\date{Institute of Mechanics,
Moscow State Lomonosov University, \\ Michurinskii pr. 1, 119192
Moscow, Russia \\ E-mail: kirillov@imec.msu.ru}

\begin{document}
\maketitle

\begin{abstract}
The paper presents a new theory of unfolding of eigenvalue
surfaces of real symmetric and Hermitian matrices due to an
arbitrary complex perturbation near a diabolic point. General
asymptotic formulae describing deformations of a conical surface
for different kinds of perturbing matrices are derived. As a
physical application, singularities of the surfaces of refractive
indices in crystal optics are studied.
\end{abstract}

\section{Introduction}

Since the papers by \cite{Neumann} and \cite{Teller} it is known
that the energy surfaces in quantum physics may cross forming two
sheets of a double cone: a diabolo. The apex of the cone is called
a diabolic point, see \cite{Berry1}. This kind of crossing is
typical for systems described by real symmetric Hamiltonians with
at least two parameters and Hermitian Hamiltonians depending on
three or more parameters. From mathematical point of view the
energy surfaces are described by eigenvalues of real symmetric or
Hermitian operators dependent on parameters, and the diabolic
point is a point of a double eigenvalue with two linearly
independent eigenvectors. In modern problems of quantum physics,
crystal optics, physical chemistry, acoustics and mechanics it is
important to know how the diabolic point bifurcates under
arbitrary complex perturbations forming topological singularities
of eigenvalue surfaces like a double coffee filter with two
exceptional points or a diabolic circle of exceptional points, see
e.g. \cite{MH}, \cite{MH1}, \cite{SS}, \cite{KKM}, \cite{Berry2},
\cite{KKM1}, \cite{Berry}.

     In our preceding companion paper \cite{MKS}, a general theory of coupling of
     eigenvalues for complex matrices of arbitrary dimension smoothly
     depending on multiple real parameters was presented. Two kinds of
     important singularities were mathematically classified:
     the diabolic points (DPs) and the exceptional points (EPs).
     DP is a point where the eigenvalues coalesce, while corresponding
     eigenvectors remain different (linearly independent), and EP is
     a point where both eigenvalues and eigenvectors coalesce
     forming a Jordan block. General formulae describing coupling
     and decoupling of eigenvalues, crossing and
     avoided crossing of eigenvalue surfaces were derived.
     Both the DP and EP cases are interesting
     in applications and were observed in experiments, see \cite{RR},
     \cite{Dembowsky}, \cite{Dembowsky2003}, \cite{Stehmann}.

     In the present paper following the theory developed
     in \cite{MKS} we study effects of complex perturbations
     in multiparameter families of real symmetric and Hermitian
     matrices. In case of real symmetric matrices we study unfolding
     of eigenvalue surfaces near a diabolic point under real and complex
     perturbations. Origination of singularities like a "double coffee
     filter"
     and a "diabolic circle" is analytically described.
     Unfolding of a diabolic point of a Hermitian matrix under an
     arbitrary complex perturbation is analytically treated.
     We emphasize that the unfolding of eigenvalue surfaces is described
     qualitatively as well as quantitatively by using only
     the information at the diabolic point,
     including eigenvalues, eigenvectors, and derivatives of the system matrix
     taken at the diabolic point.

     As a physical application, singularities of the surfaces of refractive
     indices in crystal optics are studied. Asymptotic formulae
     for the metamorphoses of these surfaces depending on properties of a crystal
     are established and discussed in detail. Singular axes for general crystals
     with weak absorption and chirality are found. A new explicit
     condition distinguishing the absorbtion-dominated and
     chirality-dominated crystals is established in terms of components of
     the inverse dielectric tensor.

\section{Asymptotic expressions for eigenvalues near a diabolic point}

Let us consider the eigenvalue problem
    \begin{equation}
    \mathbf{A}\mathbf{u} = \lambda\mathbf{u}
    \label{eq0.1}
    \end{equation}
for an $m\times m$ Hermitian matrix $\mathbf{A}$, where $\lambda$
is an eigenvalue and $\mathbf{u}$ is an eigenvector. Such
eigenvalue problems arise in non-dissipative physics with and
without time reversal symmetry. Real symmetric and complex
Hermitian matrices correspond to these two cases, respectively. We
assume that the matrix $\mathbf{A}$ smoothly depends on a vector
of $n$ real parameters $\mathbf{p} = (p_1,\ldots,p_n)$. Let
$\lambda_0$ be a double eigenvalue of the matrix $\mathbf{A}_0 =
\mathbf{A}(\mathbf{p}_0)$ for some vector ${\bf p}_0$. Since
$\mathbf{A}_0$ is a Hermitian matrix, the eigenvalue $\lambda_0$
is real and possesses two eigenvectors $\mathbf{u}_1$ and
$\mathbf{u}_2$. Thus, the point of eigenvalue coupling for
Hermitian matrices is diabolic. We choose the eigenvectors
satisfying the normalization conditions
    \begin{equation}
    (\mathbf{u}_1,\mathbf{u}_1)
    = (\mathbf{u}_2,\mathbf{u}_2) = 1,\quad
    (\mathbf{u}_1,\mathbf{u}_2) = 0,
    \label{eq0.1b}
    \end{equation}
where the standard inner product of complex vectors is given by
$(\mathbf{u},\mathbf{v}) = \sum_{i = 1}^m u_i\overline{v}_i$.

Under perturbation of parameters $\mathbf{p} =
\mathbf{p}_0+\Delta\mathbf{p}$, the bifurcation of $\lambda_0$
into two simple eigenvalues $\lambda_+$ and $\lambda_-$ occurs.
The asymptotic formula for $\lambda_\pm$ under multiparameter
perturbation is \cite{MKS}
    \begin{equation}
    \lambda_\pm
    = \lambda_0+\frac{\langle\mathbf{f}_{11}+\mathbf{f}_{22},
    \Delta\mathbf{p}\rangle}{2}
    \pm\sqrt{\frac{\langle\mathbf{f}_{11}-\mathbf{f}_{22},
    \Delta\mathbf{p}\rangle^2}{4}
    +\langle\mathbf{f}_{12},\Delta\mathbf{p}\rangle
    \langle\mathbf{f}_{21},\Delta\mathbf{p}\rangle}.
    \label{eq0.2}
    \end{equation}
Components of the vector $\mathbf{f}_{ij} =
(f_{ij}^1,\ldots,f_{ij}^n)$ are
    \begin{equation}
    f_{ij}^k = \left(\frac{\partial\mathbf{A}}{\partial p_k}
    \mathbf{u}_i,\mathbf{u}_j\right),
    \label{eq0.4}
    \end{equation}
where the derivative is taken at $\mathbf{p}_0$, and inner
products of vectors in (\ref{eq0.2}) are given by
$\langle\mathbf{a},\mathbf{b}\rangle = \sum_{i = 1}^n
a_i\overline{b}_i$. In expression (\ref{eq0.2}) the higher order
terms $o(\|\Delta\mathbf{p}\|)$ and $o(\|\Delta\mathbf{p}\|^2)$
are neglected before and under the square root. Since the matrix
$\mathbf{A}$ is Hermitian, the vectors $\mathbf{f}_{11}$ and
$\mathbf{f}_{22}$ are real and the vectors $\mathbf{f}_{12} =
\overline{\mathbf{f}}_{21}$ are complex conjugate. In case of real
symmetric matrices $\mathbf{A} = \mathbf{A}^T$, the vectors
$\mathbf{f}_{11}$, $\mathbf{f}_{22}$, and $\mathbf{f}_{12} =
\mathbf{f}_{21}$ are real. The asymptotic expression for the
eigenvectors corresponding to $\lambda_\pm$ takes the form
\cite{MKS}
    \begin{equation}
    \mathbf{u}_\pm
    = \alpha_{\pm}\mathbf{u}_1+\beta_{\pm}\mathbf{u}_2,\quad
    \frac{\alpha_{\pm}}{\beta_{\pm}}
    = \frac{\langle\mathbf{f}_{12},\Delta\mathbf{p}\rangle}{
    \lambda_\pm-\lambda_0
    -\langle\mathbf{f}_{11},\Delta\mathbf{p}\rangle}
    =\frac{\lambda_\pm-\lambda_0-\langle\mathbf{f}_{22},\Delta\mathbf{p}
    \rangle}{\langle\mathbf{f}_{21},\Delta\mathbf{p}\rangle}.
    \label{eq0.5}
    \end{equation}
Expressions (\ref{eq0.5}) provide zero order terms for the
eigenvectors $\mathbf{u}_\pm$ under perturbation of the parameter
vector.

Now, consider an arbitrary complex perturbation of the matrix
family $\mathbf{A}(\mathbf{p})+\Delta\mathbf{A}(\mathbf{p})$. Such
perturbations appear due to non-conservative effects breaking
symmetry of the initial system. We assume that the size of
perturbation $\Delta\mathbf{A}(\mathbf{p}) \sim \varepsilon$ is
small, where $\varepsilon = \|\Delta\mathbf{A}(\mathbf{p}_0)\|$ is
the Frobenius norm of the perturbation at the diabolic point.
Behavior of the eigenvalues $\lambda_\pm$ for small
$\Delta\mathbf{p}$ and small $\varepsilon$ is described by the
following asymptotic formula~\cite{MKS}
    \begin{equation}
    \begin{array}{rcl}
        \lambda_\pm & = & \displaystyle
        \lambda_0+\frac{\langle\mathbf{f}_{11}+\mathbf{f}_{22},
        \Delta\mathbf{p}\rangle}{2}
        +\frac{\varepsilon_{11}+\varepsilon_{22}}{2} \\[15pt]
        & & \displaystyle
        \pm\,\sqrt{\frac{(\langle\mathbf{f}_{11}-\mathbf{f}_{22},
        \Delta\mathbf{p}\rangle+\varepsilon_{11}-\varepsilon_{22})^2}{4}
        +(\langle\mathbf{f}_{12},\Delta\mathbf{p}\rangle
        +\varepsilon_{12})
        (\langle\mathbf{f}_{21},\Delta\mathbf{p}\rangle
        +\varepsilon_{21})}.
    \end{array}
    \label{eq0.6}
    \end{equation}
The quantities $\varepsilon_{ij}$ are small complex numbers of
order $\varepsilon$ given by the expression
    \begin{equation}
    \varepsilon_{ij}
    =
    \left(\Delta\mathbf{A}(\mathbf{p}_0)\mathbf{u}_i,\mathbf{u}_j\right).
    \label{eq0.8}
    \end{equation}
A small variation of the matrix family leads to the following
correction of the asymptotic expression for the eigenvectors
    \begin{equation}
    \mathbf{u}_\pm
    = \alpha^\varepsilon_{\pm}\mathbf{u}_1
    +\beta^\varepsilon_{\pm}\mathbf{u}_2,\quad
    \frac{\alpha^\varepsilon_{\pm}}{\beta^\varepsilon_{\pm}}
    = \frac{\langle\mathbf{f}_{12},\Delta\mathbf{p}\rangle
    +\varepsilon_{12}}{\lambda_\pm-\lambda_0
    -\langle\mathbf{f}_{11},\Delta\mathbf{p}\rangle
    -\varepsilon_{11}}
    =\frac{\lambda_\pm-\lambda_0-\langle\mathbf{f}_{22},\Delta\mathbf{p}
    \rangle-\varepsilon_{22}}{
    \langle\mathbf{f}_{21},\Delta\mathbf{p}\rangle
    +\varepsilon_{21}}.
    \label{eq0.9}
    \end{equation}
The ratios $\alpha^\varepsilon_+/\beta^\varepsilon_+ =
\alpha^\varepsilon_-/\beta^\varepsilon_-$ at the point of
coincident eigenvalues $\lambda_+ = \lambda_-$. Hence, the
eigenvectors $\mathbf{u}_+ = \mathbf{u}_-$ coincide, and the point
of eigenvalue coupling of the perturbed system becomes exceptional
(EP). For some specific perturbations
$\Delta\mathbf{A}(\mathbf{p})$, the coupling point may remain
diabolic under the conditions
\begin{equation}
    \langle\mathbf{f}_{12},\Delta\mathbf{p}\rangle
    +\varepsilon_{12}=0,~~
    \langle\mathbf{f}_{21},\Delta\mathbf{p}\rangle
    +\varepsilon_{21}=0,~~
    \langle\mathbf{f}_{11}-\mathbf{f}_{22},\Delta\mathbf{p}\rangle
    +\varepsilon_{11}-\varepsilon_{22}=0,
    \label{eq0.10}
\end{equation}
when both ratios in (\ref{eq0.9}) become undetermined.

We observe that asymptotic description of unfolding of diabolic
singularity due to perturbation of the matrix family requires only
the value of $\Delta\mathbf{A}(\mathbf{p})$ taken at the coupling
point $\mathbf{p}_0$. Dependence of the perturbation
$\Delta\mathbf{A}$ on the vector of parameters $\mathbf{p}$ near
the point $\mathbf{p}_0$ is not so important, since it influences
higher order terms.

\section{Unfolding of a diabolic singularity for real symmetric matrices}

Let us assume that ${\bf A}({\bf p})$ is an $n$-parameter family
of real symmetric matrices. Then its eigenvalues $\lambda$ are
real. Let $\lambda_0$ be a double eigenvalue of the matrix
$\mathbf{A}_0 = \mathbf{A}(\mathbf{p}_0)$ with two real
eigenvectors $\mathbf{u}_1$ and $\mathbf{u}_2$. Under perturbation
of parameters $\mathbf{p} = \mathbf{p}_0+\Delta\mathbf{p}$, the
eigenvalue $\lambda_0$ splits into two simple eigenvalues
$\lambda_+$ and $\lambda_-$. The asymptotic formula for
$\lambda_\pm$ under multiparameter perturbation is given by
equations (\ref{eq0.2}) and (\ref{eq0.4}), where the vectors ${\bf
f}_{11}$, ${\bf f}_{22}$, and ${\bf f}_{12}={\bf f}_{21}$ are
real. Then, equation (\ref{eq0.2}) takes the form
    \begin{equation}
    \left(\lambda_{\pm}-\lambda_0-\frac{\langle {\bf f}_{11}+{\bf f}_{22},\Delta
    {\bf p}\rangle}{2}\right)^2-\frac{\langle {\bf f}_{11}-{\bf f}_{22},\Delta
    {\bf p}\rangle^2}{4}-\langle {\bf f}_{12},\Delta
    {\bf p}\rangle^2=0.
    \label{eq1.1}
    \end{equation}

Equation (\ref{eq1.1}) describes a surface in the space
$(p_1,p_2,\ldots,p_n,\lambda)$, which consists of two sheets
$\lambda_+(\bf p)$ and $\lambda_-(\bf p)$. These sheets are
connected at the points satisfying the equations
    \begin{equation}
    \lambda_{\pm}=\lambda_0+\frac{1}{2}\langle {\bf f}_{11}+{\bf
    f}_{22},\Delta{\bf p}\rangle,~~
    \langle {\bf f}_{11}-{\bf f}_{22},\Delta
    {\bf p}\rangle=0,~~\langle {\bf f}_{12},\Delta
    {\bf p}\rangle=0,
    \label{eq1.2}
    \end{equation}
where the eigenvalues couple: $\lambda_+ = \lambda_-$. Equations
(\ref{eq1.2}) define a plane of dimension $n-2$. Thus, the double
eigenvalue is a phenomenon of codimension 2 in an $n$-parameter
family of real symmetric matrices \cite{Neumann}.

\begin{figure}
    \begin{center}
    \includegraphics[angle=0, width=0.55\textwidth]{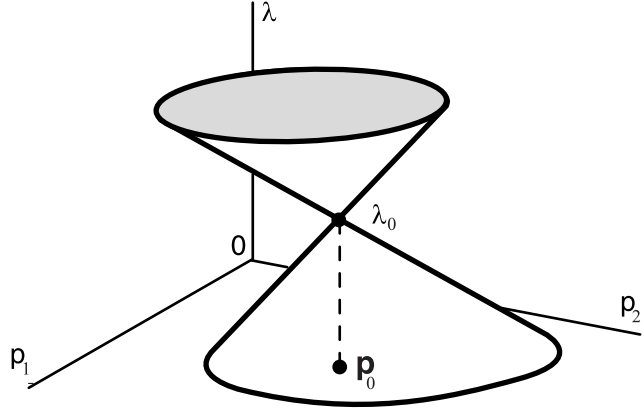}
    \end{center}
    \caption{A diabolic point in a family of real symmetric matrices.}
    \label{fig1}
    \end{figure}

For the two-parameter matrix ${\bf A}(p_1,p_2)$ equation
(\ref{eq1.1}) defines a double cone with apex at the point $({\bf
p}_0, \lambda_0)$ in the space $(p_1,p_2,\lambda)$, see
Figure~\ref{fig1}. The point $({\bf p}_0,\lambda_0)$ is referred
to as a "diabolic point" \cite{Berry1} due to the conical shape of
the children's toy "diabolo".

Let us consider a perturbation ${\bf A}({\bf p})+\Delta{\bf
A}({\bf p})$ of the real symmetric family ${\bf A}({\bf p})$ in
the vicinity of the diabolic point ${\bf p}_0$, where $\Delta {\bf
A}({\bf p})$ is a complex matrix with the small norm
$\varepsilon=\|\Delta{\bf A}({\bf p}_0) \|$. Splitting of the
double eigenvalue $\lambda_0$ due to a change of the vector of
parameters $\Delta {\bf p}$ and a small complex perturbation
$\Delta{\bf A}$ is described by equation (\ref{eq0.6}), which
acquires the form
    \begin{equation}
    \lambda_{\pm}=\lambda'_0+\mu \pm \sqrt{c},~~
    c=(x+\xi)^2+(y+
    \eta)^2-\zeta^2.
    \label{eq1.3}
    \end{equation}
In equation (\ref{eq1.3}) the quantities $\lambda'_0$, $x$, and
$y$ are real:
    \begin{equation}
    \lambda'_0=\lambda_0+\frac{1}{2}\langle {\bf f}_{11}+{\bf
    f}_{22},\Delta{\bf p}\rangle,~~
    x=\frac{1}{2}\langle {\bf f}_{11}-{\bf f}_{22},\Delta
    {\bf p}\rangle,~~y=\langle {\bf f}_{12},\Delta
    {\bf p}\rangle,
    \label{eq1.4}
    \end{equation}
while the small coefficients $\mu$, $\xi$, $\eta$, and $\zeta$ are
complex:
    \begin{equation}
    \mu=\frac{1}{2}({\varepsilon}_{11}+{\varepsilon}_{22}),~~
    \xi=\frac{1}{2}({\varepsilon}_{11}-{\varepsilon}_{22}),~~
    \eta=\frac{1}{2}({\varepsilon}_{12}+{\varepsilon}_{21}),~~
    \zeta=\frac{1}{2}({\varepsilon}_{12}-{\varepsilon}_{21}).
    \label{eq1.5}
    \end{equation}
Separating real and imaginary parts in equation (\ref{eq1.3}), we
find
    \begin{equation}
    {\rm Re}^2(\lambda-\lambda'_0-\mu)-{\rm
    Im}^2(\lambda-\lambda'_0-\mu)={\rm Re}c,~~
    2{\rm Re}(\lambda-\lambda'_0-\mu){\rm
    Im}(\lambda-\lambda'_0-\mu)={\rm Im}c,
    \label{eq1.6}
    \end{equation}
where
    \begin{equation}
    {\rm Re}c=({\rm Im}^2\zeta{-}{\rm Im}^2\xi{-}{\rm Im}^2\eta{-}{\rm
    Re}^2\zeta){+}
    (x{+}{\rm Re}\xi)^2{+}(y{+}{\rm Re}\eta)^2,
    \label{eq1.7}
    \end{equation}
    \begin{equation}
    {\rm Im}c=2((x{+}{\rm Re}\xi){\rm Im}\xi+
    (y{+}{\rm Re}\eta){\rm Im}\eta-{\rm
    Re}\zeta{\rm Im}\zeta).
    \label{eq1.8}
    \end{equation}

From equations (\ref{eq1.6}) we get the expressions determining
the real and imaginary parts of the perturbed eigenvalues
    \begin{equation}
    {\rm Re}\lambda_{\pm}=\lambda'_0+{\rm Re}\mu\pm\sqrt{\left({\rm
    Re}c+\sqrt{{\rm Re}^2c+{\rm Im}^2c}~\right)/2},
    \label{eq1.9}
    \end{equation}
    \begin{equation}
    {\rm Im}\lambda_{\pm}={\rm Im}\mu\pm \sqrt{\left(-{\rm
    Re}c+\sqrt{{\rm Re}^2c+{\rm Im}^2c}~\right)/2}.
    \label{eq1.10}
    \end{equation}
Strictly speaking, for the same eigenvalue one should take equal
or opposite signs before the square roots in (\ref{eq1.9}),
(\ref{eq1.10}) for positive or negative ${\rm Im}c$, respectively.

Equations (\ref{eq1.9}) and (\ref{eq1.10}) define surfaces in the
spaces $(p_1,p_2,\ldots,p_n,{\rm Re}\lambda)$ and
$(p_1,p_2,\ldots,p_n,{\rm Im}\lambda)$. Two sheets of the surface
(\ref{eq1.9}) are connected (${\rm Re}\lambda_+={\rm
Re}\lambda_-$) at the points satisfying the conditions
\begin{equation}
    {\rm Re}c\le0,~~ {\rm Im}c=0,~~{\rm
    Re}\lambda_{\pm}=\lambda'_0+{\rm Re}\mu,
    \label{eq1.11}
\end{equation}
while the sheets ${\rm Im}\lambda_+({\bf p})$ and ${\rm
Im}\lambda_-({\bf p})$ are glued at the set of points satisfying
\begin{equation}
    {\rm Re}c\ge0,~~ {\rm Im}c=0,~~{\rm
    Im}\lambda_{\pm}={\rm Im}\mu.
    \label{eq1.12}
\end{equation}

Note that in the neighborhood of the intersections (\ref{eq1.11})
and (\ref{eq1.12}) the eigenvalue sheets given by the formulae
(\ref{eq1.9}) and (\ref{eq1.10}) can be described by the following
approximate expressions
    \begin{equation}
    {\rm Re}\lambda_{\pm}{=}\lambda'_0{+}{\rm Re}\mu\pm\frac{{\rm Im}c}{2}\sqrt{\frac{-1}{{\rm
    Re}c}},~~{\rm Re}c<0;~~    {\rm Im}\lambda_{\pm}{=}{\rm Im}\mu\pm\frac{{\rm Im}c}{2}\sqrt{\frac{1}{{\rm
    Re}c}},~~{\rm Re}c>0.
    \label{eq1.13}
    \end{equation}

The eigenvalue remains double under the perturbation of parameters
when $c = 0$, which yields two equations ${\rm Re}c = 0$ and ${\rm
Im}c = 0$. Two cases are distinguished according to the sign of
the quantity
\begin{equation}
    D={\rm Im}^2\xi{+}{\rm Im}^2\eta{-}{\rm Im}^2\zeta.
    \label{eq1.17a}
\end{equation}
If $D > 0$, then the equations ${\rm Re}c = 0$ and ${\rm Im}c = 0$
with expressions (\ref{eq1.7}), (\ref{eq1.8}) yield two solutions
$(x_a,y_a)$ and $(x_b,y_b)$, where
    \begin{equation}
    x_{a,b}=\frac{{\rm Im}\xi{\rm Re}\zeta{\rm Im}\zeta \pm
    {\rm Im}\eta\sqrt{({\rm Im}^2\xi{+}{\rm Im}^2\eta{+}{\rm Re}^2\zeta)
    ({\rm Im}^2\xi{+}{\rm Im}^2\eta{-}{\rm Im}^2\zeta)}}
    {{\rm Im}^2\xi{+}{\rm Im}^2\eta}-{\rm Re}\xi,
    \label{eq1.14}
    \end{equation}
    \begin{equation}
    y_{a,b}=\frac{{\rm Im}\eta{\rm Re}\zeta{\rm Im}\zeta \mp
    {\rm Im}\xi\sqrt{({\rm Im}^2\xi{+}{\rm Im}^2\eta{+}{\rm Re}^2\zeta)
    ({\rm Im}^2\xi{+}{\rm Im}^2\eta{-}{\rm Im}^2\zeta)}}
    {{\rm Im}^2\xi{+}{\rm Im}^2\eta}-{\rm Re}\eta.
    \label{eq1.15}
    \end{equation}
These two solutions determine the points in parameter space, where
double eigenvalues appear. When $D = 0$, the two solutions
coincide. For $D < 0$, the equations ${\rm Re}c = 0$ and ${\rm
Im}c = 0$ have no real solutions. In the latter case, the
eigenvalues $\lambda_+$ and $\lambda_-$ separate for all $\Delta
{\bf p}$.

    \begin{figure}
    \begin{center}
    \includegraphics[angle=0, width=0.73\textwidth]{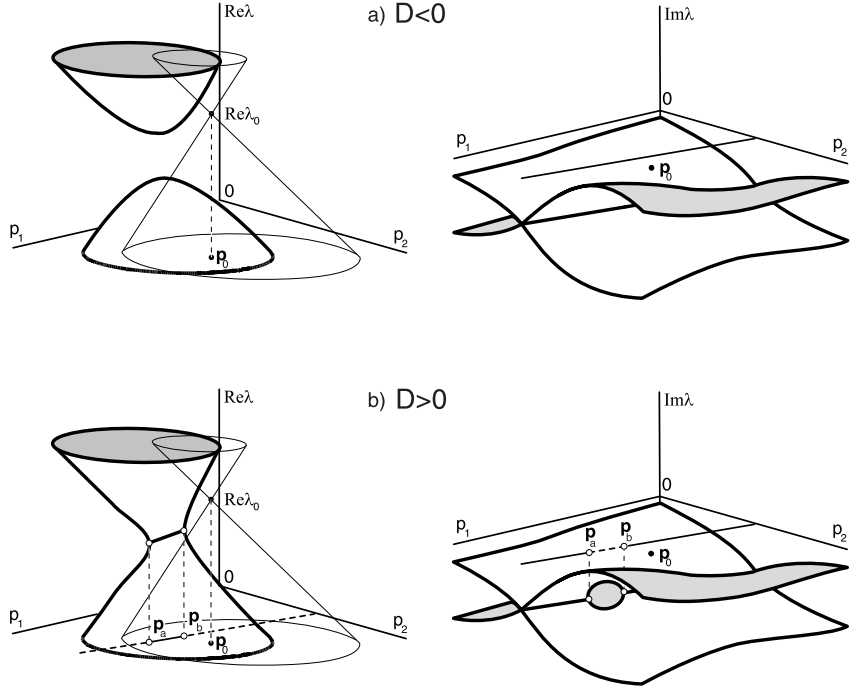}
    \end{center}
    \caption{Unfolding of a diabolic point due to complex perturbation.}
    \label{fig2}
    \end{figure}

Note that the quantities ${\rm Im}\xi$ and ${\rm Im}\eta$ are
expressed by means of the anti-Hermitian part $\Delta{\bf
A}_N=(\Delta{\bf A}-\overline{\Delta{\bf A}}^T)/2$ of the matrix
$\Delta{\bf A}$ as
    \begin{equation}
    \begin{array}{l}
        \displaystyle
        {\rm Im}\xi
        = \frac{(\Delta{\bf A}_N({\bf p}_0){\bf u}_1,{\bf u}_1)
        {-}(\Delta{\bf A}_N({\bf p}_0){\bf u}_2,{\bf
        u}_2)}{2i},\\[10pt]
        \displaystyle
        {\rm Im}\eta
        = \frac{(\Delta{\bf A}_N({\bf p}_0){\bf u}_1,{\bf u}_2)
        {+}(\Delta{\bf A}_N({\bf p}_0){\bf u}_2,{\bf u}_1)}{2i},
    \end{array}
    \label{eq1.16}
    \end{equation}
while ${\rm Im}\zeta$ depends on the Hermitian part $\Delta{\bf
A}_H=(\Delta{\bf A}+\overline{\Delta{\bf A}}^T)/2$ as
    \begin{equation}
    {\rm Im}\zeta
    = \frac{(\Delta{\bf A}_H({\bf p}_0){\bf u}_1,{\bf u}_2)
    -(\Delta{\bf A}_H({\bf p}_0){\bf u}_2,{\bf u}_1)}{2i}.
    \label{eq1.17}
    \end{equation}
If $D>0$, one can say that the influence of the anti-Hermitian
part of the perturbation $\Delta{\bf A}$ is stronger than that of
the Hermitian part. If the Hermitian part prevails in the
perturbation $\Delta{\bf A}$, we have $D < 0$. In particular, $D =
{-}{\rm Im}^2\zeta < 0$ for a purely Hermitian perturbation
$\Delta {\bf A}$.

Let us assume that the vector ${\bf p}$ consists of only two
components $p_1$ and $p_2$, and consider the surfaces
(\ref{eq1.9}) and (\ref{eq1.10}) for different kinds of the
perturbation $\Delta{\bf A}(\bf p)$. Consider first the case
$D<0$. Then, the eigensheets ${\rm Re}\lambda_+({\bf p})$ and
${\rm Re}\lambda_-({\bf p})$ are separate, see Figure~\ref{fig2}a.
Indeed, for $D\le -{\rm Re}^2\zeta$ the inequality ${\rm Re}c\ge0$
holds for all variations of parameters, see equation
(\ref{eq1.7}). In the case when $-{\rm Re}^2\zeta<D<0$ the
equation ${\rm Re}c=0$ with expressions (\ref{eq1.4}) define an
ellipse in the plane of parameters $(p_1,p_2)$. Inside the ellipse
we have ${\rm Re}c<0$ and outside ${\rm Re}c>0$. Equation ${\rm
Im}c=0$ defines a line in parameter plane. The line and the
ellipse have no common points for $D<0$ since there are no real
solutions of the equation $c=0$. Hence, for $D<0$ conditions
(\ref{eq1.11}) are not fulfilled and the real parts of the
eigenvalues avoid crossing. As the size of the complex
perturbation decreases ($\varepsilon \rightarrow 0$), the two
sheets come closer and touch each other at the point $({\bf
p}_0,\lambda_0)$ for $\varepsilon=0$ forming the diabolic
singularity. The sheets ${\rm Im}\lambda_+({\bf p})$ and ${\rm
Im}\lambda_-({\bf p})$ of the eigensurface (\ref{eq1.10})
intersect along the line
    \begin{equation}
    {\rm Im}c/2 = (x{+}{\rm Re}\xi){\rm Im}\xi+
    (y{+}{\rm Re}\eta){\rm Im}\eta-{\rm
    Re}\zeta{\rm Im}\zeta=0,~~{\rm
    Im}\lambda_{\pm}={\rm Im}\mu,
    \label{eq1.18}
    \end{equation}
given by conditions (\ref{eq1.12}). Note that, by using
(\ref{eq1.13}), one can show that the angle of intersection of the
imaginary eigensheets is small of order $\varepsilon$ and tends to
zero as $\varepsilon \rightarrow 0$.

In the case $D>0$ the line ${\rm Im}c=0$ and the ellipse ${\rm
Re}c=0$ have common points ${\bf p}_{a}$ and ${\bf p}_{b}$ where
the eigenvalues couple. Coordinates of these points found from the
equations (\ref{eq1.4}) are
    \begin{equation}
    {\bf p}_{a,b}{=}{\bf p}_0{+}\left({-}\frac{2f_{12}^2x_{a,b}{-}(f_{11}^2{-}f_{22}^2)y_{a,b}}
    {f_{12}^1(f_{11}^2{-}f_{22}^2){-}f_{12}^2(f_{11}^1{-}f_{22}^1)},~~
    \frac{2f_{12}^1x_{a,b}{-}(f_{11}^1{-}f_{22}^1)y_{a,b}}
    {f_{12}^1(f_{11}^2{-}f_{22}^2){-}f_{12}^2(f_{11}^1{-}f_{22}^1)}\right),
    \label{eq1.19}
    \end{equation}
where $x_{a,b}$ and $y_{a,b}$ are defined by expressions
(\ref{eq1.14}) and (\ref{eq1.15}). Here we have assumed that the
vectors ${\bf f}_{11}-{\bf f}_{22}$ and ${\bf f}_{12}$ are
linearly independent. Note that the points ${\bf p}_a$ and ${\bf
p}_b$ coincide in the degenerate case $D=0$.

According to conditions (\ref{eq1.11}) the real eigensheets ${\rm
Re}\lambda_+({\bf p})$ and ${\rm Re}\lambda_-({\bf p})$ are glued
in the interval $[{\bf p}_a, {\bf p}_b]$ of the line
    \begin{equation}
    {\rm Im}c/2=(x{+}{\rm Re}\xi){\rm Im}\xi+
    (y{+}{\rm Re}\eta){\rm Im}\eta-{\rm
    Re}\zeta{\rm Im}\zeta=0,~~
    {\rm Re}\lambda_{\pm}=\lambda'_0+{\rm Re}\mu.
    \label{eq1.20}
    \end{equation}
The surface of real eigenvalues (\ref{eq1.9}) is called a "double
coffee filter" \cite{KKM}. The unfolding of a diabolic point into
the double coffee filter is shown in Figure~\ref{fig2}b.

From conditions (\ref{eq1.12}) it follows that the imaginary
eigensheets ${\rm Im}\lambda_+({\bf p})$ and ${\rm
Im}\lambda_-({\bf p})$ are connected along the straight line
(\ref{eq1.18}) where the interval $[{\bf p}_a, {\bf p}_b]$ is
excluded, see Figure~\ref{fig2}b. According to the formulae
(\ref{eq1.13}) the angle of intersection of the imaginary
eigensheets tends to $\pi$ as the points ${\bf p}_a$ and ${\bf
p}_b$ are approached, since ${\rm Re}c$ goes to zero. At far
distances from the interval $[{\bf p}_a, {\bf p}_b]$ this angle
becomes small of order $\varepsilon$. With the decrease of the
size of complex perturbation $\varepsilon$ the interval shrinks
and the angle of intersection goes to zero. At $\varepsilon=0$ the
imaginary parts of the eigenvalues coincide: ${\rm
Im}\lambda_+={\rm Im}\lambda_-=0$. Note that in crystal optics and
acoustics the interval $[{\bf p}_a, {\bf p}_b]$ is referred to as
a "branch cut", and the points ${\bf p}_a$, ${\bf p}_b$ are called
"singular axes", see \cite{SS, Berry2, RR}. According to equation
(\ref{eq0.9}) the double eigenvalues at ${\bf p}_a$ and ${\bf
p}_b$ possess only one eigenvector and, hence, they are
exceptional points (EPs).

    \begin{figure}
        \begin{center}
        \includegraphics[angle=0, width=0.9\textwidth]{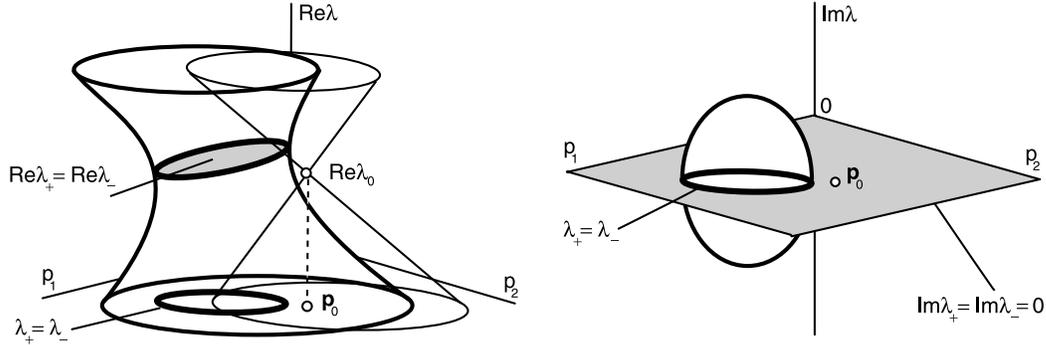}
        \end{center}
    \caption{A real non-symmetric perturbation of a diabolic point.}
    \label{fig3}
    \end{figure}

Now let us consider the case when the perturbation $\Delta {\bf
A}({\bf p})$ is real. In this case $\mu$, $\xi$, $\eta$, $\zeta$,
and hence,
    \begin{equation}
    c=(x+\xi)^2+(y+\eta)^2-\zeta^2
    \label{eq1.21}
    \end{equation}
are real quantities. According to (\ref{eq1.3}) the eigenvalues
$\lambda_{\pm}$ are complex-conjugate if $c<0$ and real if $c>0$.
The eigenvalues couple for $c=0$ forming a set consisting of
exceptional points with double real eigenvalues.

Consider a system depending on a vector of two parameters ${\bf
p}=(p_1,p_2)$. Then equation $c=0$ with expressions (\ref{eq1.4})
and (\ref{eq1.21}) define an ellipse in parameter plane; $c<0$
inside the ellipse and $c>0$ outside. Real parts of the
eigenvalues are given by the equations
    \begin{eqnarray}
    c \ge 0: && ({\rm Re}\lambda{-}\lambda'_0-\mu)^2{-}
    (x+\xi)^2{-}(y+\eta)^2{=}-\zeta^2,
    \label{eq1.22} \\
     c \le 0: && {\rm Re}\lambda=\lambda'_0+\mu.
    \label{eq1.23}
    \end{eqnarray}
Equation (\ref{eq1.22}) defines a hyperboloid in the space
$(p_1,p_2,{\rm Re}\lambda)$. Real parts of the eigenvalues
$\lambda_{\pm}$ coincide at the disk determined by equation
(\ref{eq1.23}), see Figure~\ref{fig3}. Imaginary parts of the
eigenvalues are
    \begin{eqnarray}
    c \ge 0: && {\rm Im}\lambda=0,
    \label{eq1.24} \\
     c \le 0: && {\rm Im}^2\lambda+
    (x+\xi)^2{+}(y+\eta)^2=\zeta^2.
    \label{eq1.25}
    \end{eqnarray}
The imaginary parts are both zero at the points of the plane
(\ref{eq1.24}) surrounding the ellipsoid (\ref{eq1.25}) ("a
bubble") in the space $(p_1,p_2,{\rm Im}\lambda)$, see
Figure~\ref{fig3}. The eigenvalues couple at the points of the
elliptic ring
    \begin{equation}
    \lambda_{\pm}=\lambda'_0+\mu,~~(x+\xi)^2{+}(y+\eta)^2=\zeta^2,
    \label{eq1.26}
    \end{equation}
consisting of exceptional points, see Figure~\ref{fig3}. By that
reason we call it an "exceptional ring", which is a better name
compared with a "diabolic circle" suggested by \cite{MH},
\cite{MH1}.

Finally, it is instructive to consider deformations of the
surfaces (\ref{eq1.22}), (\ref{eq1.23}) and (\ref{eq1.24}),
(\ref{eq1.25}) as the real perturbation becomes complex. If the
imaginary part of the perturbation ${\rm Im}\Delta{\bf A}$ is such
that $D<0$, then the parts of the hyperboloid (\ref{eq1.22})
connected by the disk (\ref{eq1.23}) are separated into the two
smooth surfaces described by the equation (\ref{eq1.9}). On the
other hand, the ellipsoid (\ref{eq1.25}) surrounded by the plane
(\ref{eq1.24}) is foliated into two sheets crossing each other
along the line ${\rm Im}c=0$, see Figure~2a. Recall that the line
${\rm Im}c=0$ does not intersect the ellipse ${\rm Re}c=0$. When
$D>0$, the disk (\ref{eq1.23}) foliates into two sheets crossing
along the interval $[{\bf p}_a,{\bf p}_b]$, where the points ${\bf
p}_a$ and ${\bf p}_b$ are given by expression (\ref{eq1.19}). As
the size of the imaginary part of the perturbation ${\rm
Im}\Delta{\bf A}$ increases, the angle of intersection of real
eigensheets grows. By this way, the purely imaginary perturbation
deforms the hyperboloid (\ref{eq1.22}) into the double coffee
filter (\ref{eq1.9}), see Figure~\ref{fig2}b. The ellipsoid
(\ref{eq1.25}) surrounded by the plane (\ref{eq1.24}) is
transformed into two smooth sheets intersecting along the line
${\rm Im}c=0$, where the interval $[{\bf p}_a,{\bf p}_b]$ is
excluded. The angle of intersection grows as the size of the
perturbation ${\rm Im}\Delta{\bf A}$ increases.

\section{Unfolding of a diabolic singularity for Hermitian matrices}

Let us consider a multi-parameter Hermitian matrix
$\mathbf{A}(\mathbf{p})$. Assume that $\mathbf{p}_0$ is a diabolic
point, where the matrix $\mathbf{A}_0 = \mathbf{A}(\mathbf{p}_0)$
has a double real eigenvalue $\lambda_0$ with two eigenvectors.
The splitting of $\lambda_0$ into a pair of simple real
eigenvalues $\lambda_+$ and $\lambda_-$ is described by
expressions (\ref{eq0.2}), (\ref{eq0.4}), where the vectors
$\mathbf{f}_{11}$ and $\mathbf{f}_{22}$ are real and the vectors
$\mathbf{f}_{12} = \overline{\mathbf{f}}_{21}$ are complex
conjugate. By using expression (\ref{eq0.2}), we find
    \begin{equation}
    \label{Heq.1}
    \lambda_\pm = \lambda'_0\pm\sqrt{x^2+y^2+z^2},
    \end{equation}
where $\lambda'_0$, $x$, $y$, and $z$ are real quantities
depending linearly on the perturbation of parameters
$\Delta\mathbf{p}$ as follows
    \begin{equation}
    \label{Heq.2}
    \lambda'_0 = \lambda_0+\frac{\langle\mathbf{f}_{11}+\mathbf{f}_{22},
    \Delta\mathbf{p}\rangle}{2},\
    x = \frac{\langle\mathbf{f}_{11}-\mathbf{f}_{22},
    \Delta\mathbf{p}\rangle}{2},\
    y = \langle\mathrm{Re}\,\mathbf{f}_{12},
    \Delta\mathbf{p}\rangle,\
    z = \langle\mathrm{Im}\,\mathbf{f}_{12},
    \Delta\mathbf{p}\rangle.
    \end{equation}
The eigenvalues coincide if $x = y = z = 0$. Thus, if the system
depends on three parameters and the real vectors
$\mathbf{f}_{11}-\mathbf{f}_{22}$, $\mathrm{Re}\,\mathbf{f}_{12}$,
and $\mathrm{Im}\,\mathbf{f}_{12}$ are linearly independent, the
eigenvalues $\lambda_+$ and $\lambda_-$ split for any nonzero
perturbation $\Delta\mathbf{p}$. For more than three parameters,
the equations $x = y = z = 0$ with relations (\ref{Heq.2}) provide
a plane of diabolic points in parameter space. This plane has
dimension $n-3$, which agrees with the well-known fact that the
diabolic point is a codimension $3$ phenomenon for Hermitian
systems~\cite{Neumann, ArnoldQuasiMody}.

Now let us consider a general non-Hermitian perturbation of the
system $\mathbf{A}(\mathbf{p})+\Delta\mathbf{A}(\mathbf{p})$,
assuming that the size of perturbation at the diabolic point
$\varepsilon = \|\Delta\mathbf{A}(\mathbf{p}_0)\|$ is small. The
two eigenvalues $\lambda_+$ and $\lambda_-$, which become complex
due to non-Hermitian perturbation, are given by asymptotic
expressions (\ref{eq0.6}), (\ref{eq0.8}). With the use of the new
coordinates (\ref{Heq.2}), we write the expression (\ref{eq0.6})
as
    \begin{equation}
    \label{Heq.3}
    \lambda_\pm = \lambda'_0+\mu\pm\sqrt{c},
    \end{equation}
where
    \begin{equation}
    \label{Heq.4}
    c = (x+\xi)^2+(y+\eta)^2+(z-i\zeta)^2,
    \end{equation}
and $\mu$, $\xi$, $\eta$, $\zeta$ are small complex quantities of
order $\varepsilon$ given by expressions (\ref{eq1.5}).

\begin{figure}
    \begin{center}
    \includegraphics[angle=0, width=0.45\textwidth]{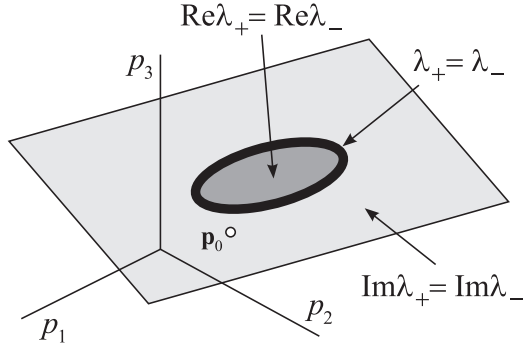}
    \end{center}
    \caption{Unfolding of a diabolic point into an
    exceptional ring in parameter space.}
    \label{Hfig.1}
    \end{figure}

The eigenvalues couple ($\lambda_+ = \lambda_-$) if $c = 0$. This
yields two equations
    \begin{equation}
    \label{Heq.5}
    \mathrm{Re}\,c =
    (x+\mathrm{Re}\,\xi)^2+(y+\mathrm{Re}\,\eta)^2
    +(z+\mathrm{Im}\,\zeta)^2
    -(\mathrm{Im}^2\xi+\mathrm{Im}^2\eta+\mathrm{Re}^2\zeta)
    = 0,
    \end{equation}
    \begin{equation}
    \label{Heq.6}
    \mathrm{Im}\,c =
    2(\mathrm{Im}\,\xi(x+\mathrm{Re}\,\xi)
    +\mathrm{Im}\,\eta(y+\mathrm{Re}\,\eta)
    -\mathrm{Re}\,\zeta(z+\mathrm{Im}\,\zeta)) = 0.
    \end{equation}
Equation (\ref{Heq.5}) defines a sphere in $(x,y,z)$ space with
the center at
$(-\mathrm{Re}\,\xi,-\mathrm{Re}\,\eta,-\mathrm{Im}\,\zeta)$ and
the radius
$\sqrt{\mathrm{Im}^2\xi+\mathrm{Im}^2\eta+\mathrm{Re}^2\zeta}$,
which are small of order $\varepsilon$. Equation (\ref{Heq.6})
yields a plane passing through the center of the sphere. The
sphere and the plane intersect along a circle. Points of this
circle determine values of parameters, for which the eigenvalues
$\lambda_\pm$ coincide. Since $c = 0$ at the coupling point,
expression (\ref{eq0.9}) for the eigenvectors takes the form
    \begin{equation}
    \label{eqH.6b}
    \mathbf{u}_\pm
    = \alpha^\varepsilon_\pm\mathbf{u}_1
    +\beta^\varepsilon_\pm\mathbf{u}_2,\quad
    \frac{\alpha^\varepsilon_\pm}{\beta^\varepsilon_\pm}
    = \frac{y+iz+\eta+\zeta}{-x-\xi}
    = \frac{x+\xi}{y-iz+\eta-\zeta}.
    \end{equation}
Thus, all points of the circle are exceptional points, where the
two eigenvectors $\mathbf{u}_-$ and $\mathbf{u}_+$ merge in
addition to the coupling of the eigenvalues $\lambda_+$ and
$\lambda_-$. By using the linear expressions (\ref{Heq.2}), the
set of exceptional points is found in the original parameter space
$\mathbf{p}$. The exceptional circle in $(x,y,z)$ space is
transformed into an exceptional elliptic ring in three-parameter
space $\mathbf{p}$, see Figure~\ref{Hfig.1}.


Let us consider the plane (\ref{Heq.6}), at which the quantity $c$
is real. By formula (\ref{Heq.3}), the real parts of the
eigenvalues $\lambda_\pm$ coincide inside the exceptional ring,
where $c < 0$, and the imaginary parts of $\lambda_\pm$ coincide
outside the exceptional ring, where $c > 0$, see the dark and
light shaded areas in Figure~\ref{Hfig.1}.

We see that, under a general complex perturbation, a diabolic
point of a three-parameter Hermitian system bifurcates into an
exceptional ring. This ring has elliptic shape and grows
proportionally to the size of perturbation $\varepsilon$. The real
and imaginary parts of the eigenvalues $\lambda_\pm$ coincide,
respectively, inside and outside the exceptional ring in the plane
of the ring.

\begin{figure}
    \begin{center}
    \includegraphics[angle=0, width=0.9\textwidth]{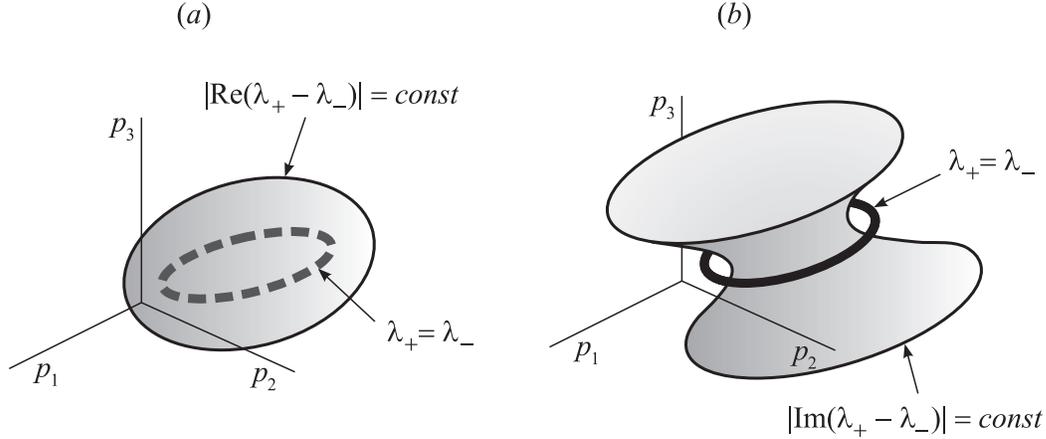}
    \end{center}
    \caption{Surfaces corresponding to coincident real or
    imaginary parts of eigenvalues.}
    \label{Hfig.2}
    \end{figure}

Finally, let us study the stratification of parameter space given
by the condition $|\mathrm{Re}\,(\lambda_+-\lambda_-)| = const$.
For problems of quantum mechanics, this difference describes the
size of a gap between two adjacent energy levels. By using
expression (\ref{Heq.3}), we find $(\lambda_+-\lambda_-)^2 = 4c$.
Separating real and imaginary parts in this equation and
extracting $\mathrm{Im}\,(\lambda_+-\lambda_-)$, we get
    \begin{equation}
    \label{Heq.7}
    \mathrm{Re}^4(\lambda_+-\lambda_-)
    -4\mathrm{Re}^2(\lambda_+-\lambda_-)\mathrm{Re}\,c
    -4\mathrm{Im}^2c = 0,
    \end{equation}
where $\mathrm{Re}\,c$ and $\mathrm{Im}\,c$ are given by the first
equalities in (\ref{Heq.5}) and (\ref{Heq.6}). Given fixed value
of $|\mathrm{Re}\,(\lambda_+-\lambda_-)|$, equation (\ref{Heq.7})
with (\ref{Heq.5}), (\ref{Heq.6}), and (\ref{Heq.2}) define an
ellipsoid in three-parameter space enclosing the exceptional ring,
see Figure~\ref{Hfig.2}a. Similar analysis provides the equation
    \begin{equation}
    \label{Heq.8}
    \mathrm{Im}^4(\lambda_+-\lambda_-)
    +4\mathrm{Im}^2(\lambda_+-\lambda_-)\mathrm{Re}\,c
    -4\mathrm{Im}^2c = 0
    \end{equation}
for a surface given by the condition
$|\mathrm{Im}\,(\lambda_+-\lambda_-)| = const$. In three-parameter
space equation (\ref{Heq.8}) defines a hyperboloid surrounded by
the exceptional ring, see Figure~\ref{Hfig.2}b.


\section{Unfolding of optical singularities of birefringent crystals}

Optical properties of a non-magnetic dichroic chiral anisotropic
crystal are characterized by the inverse dielectric tensor
$\boldsymbol\eta$, which relates the vectors of electric field
$\mathbf{E}$ and the displacement $\mathbf{D}$ as \cite{LLP}
    \begin{equation}
    \label{eqE.1}
    \mathbf{E} = {\boldsymbol \eta}\mathbf{D}.
    \end{equation}
A monochromatic plane wave of frequency $\omega$ that propagates
in a direction specified by a real unit vector $\mathbf{s} =
(s_1,s_2,s_3)$ has the form
    \begin{equation}
    \label{eqE.3}
    \mathbf{D}(\mathbf{r},t)
    = \mathbf{D}(\mathbf{s})
    \exp i\omega\!\left(\frac{n(\mathbf{s})}{c}
    \mathbf{s}^T\mathbf{r}-t\right),
    \end{equation}
where $n(\mathbf{s})$ is a refractive index, and $\mathbf{r}$ is
the real vector of spatial coordinates. With the wave
(\ref{eqE.3}) and the constitutive relation (\ref{eqE.1})
Maxwell's equations after some elementary manipulations yield (see
e.g. \cite{Berry2})
    \begin{equation}
    \label{eqE.4}
    {\boldsymbol\eta}\mathbf{D}(\mathbf{s})-\mathbf{s}
    \mathbf{s}^T{\boldsymbol\eta}\mathbf{D}(\mathbf{s})
    = \frac{1}{n^2(\mathbf{s})}\mathbf{D}(\mathbf{s}).
    \end{equation}

Multiplying equation (\ref{eqE.4}) by the vector $\mathbf{s}^T$
from the left, we find that for plane waves the vector ${\bf D}$
is always orthogonal to the direction $\bf s$, i.e., ${\bf
s}^T{\bf D}({\bf s})=0$. By using this condition, we write
(\ref{eqE.4}) in the form of the eigenvalue problem
    \begin{equation}
    \left[(\mathbf{I}-\mathbf{s}\mathbf{s}^T){\boldsymbol\eta}
    (\mathbf{I}-\mathbf{s}\mathbf{s}^T)\right]\mathbf{u}
    = \lambda\mathbf{u},
    \label{eqE.4b}
    \end{equation}
where $\lambda=n^{-2}$, ${\bf u}={\bf D}$, and ${\bf I}$ is the
identity matrix. Since $\mathbf{I}-\mathbf{s}\mathbf{s}^T$ is a
singular matrix, one of the eigenvalues is always zero. Let us
denote the other two eigenvalues by $\lambda_+$ and $\lambda_-$.
These eigenvalues determine refractive indices $n$, and the
corresponding eigenvectors yield polarizations.

The inverse dielectric tensor is described by a complex
non-Hermitian matrix $\boldsymbol\eta =
\boldsymbol\eta_{transp}+\boldsymbol\eta_{dichroic}+\boldsymbol\eta_{chiral}$.
The symmetric part of $\boldsymbol \eta$ consisting of the real
matrix $\boldsymbol\eta_{transp}$ and imaginary matrix
$\boldsymbol\eta_{dichroic}$ constitute the anisotropy tensor,
which describes the birefringence of the crystal. For a
transparent crystal, the anisotropy tensor is real and is
represented only by the matrix $\boldsymbol\eta_{transp}$; for a
crystal with linear dichroism it is complex. Choosing coordinate
axes along the principal axes of $\boldsymbol\eta_{transp}$, we
have
    \begin{equation}
    \label{eqE.2}
    {\boldsymbol\eta}_{transp} = \left(\begin{array}{ccc}
        \eta_1 & 0 & 0 \\ 0 & \eta_2 & 0 \\ 0 & 0 & \eta_3
    \end{array}\right).
    \end{equation}
The matrix
    \begin{equation}
    \label{eqE.1b}
    {\boldsymbol\eta}_{dichroic}
    = i\left(\begin{array}{ccc}
        \eta^d_{11} & \eta^d_{12} & \eta^d_{13} \\
        \eta^d_{12} & \eta^d_{22} & \eta^d_{23} \\
        \eta^d_{13} & \eta^d_{23} & \eta^d_{33}
    \end{array}\right)
    \end{equation}
describes linear dichroism (absorption). The matrix
$\boldsymbol\eta_{chiral}$ gives the antisymmetric part of
$\boldsymbol\eta$ describing chirality (optical activity) of the
crystal. It is determined by the optical activity vector
$\mathbf{g} = (g_1,g_2,g_3)$ depending linearly on $\mathbf{s}$ as
    \begin{equation}
    \label{eqE.1c}
    {\boldsymbol\eta}_{chiral}
    = i\left(\begin{array}{ccc}
        0 & -g_3 & g_2 \\
        g_3 & 0 & -g_1 \\
        -g_2 & g_1 & 0
    \end{array}\right),\quad
    \mathbf{g}
    = {\boldsymbol\gamma}\mathbf{s}
    = \left(\begin{array}{ccc}
        \gamma_{11} & \gamma_{12} & \gamma_{13} \\
        \gamma_{12} & \gamma_{22} & \gamma_{23} \\
        \gamma_{13} & \gamma_{23} & \gamma_{33}
    \end{array}\right)
    \left(\begin{array}{c}
        s_1 \\ s_2 \\ s_3
    \end{array}\right),
    \end{equation}
where $\boldsymbol\gamma$ is a symmetric optical activity tensor;
this tensor has an imaginary part for a material with circular
dichroism, see \cite{Berry2} for more details.

First, consider a transparent non-chiral crystal, when
${\boldsymbol\eta}_{dichroic} = 0$ and ${\boldsymbol\gamma} = 0$.
Then the matrix
    \begin{equation}
    \label{eqE.5}
    \mathbf{A}(\mathbf{p})
    = (\mathbf{I}-\mathbf{s}\mathbf{s}^T){\boldsymbol\eta}_{transp}
    (\mathbf{I}-\mathbf{s}\mathbf{s}^T)
    \end{equation}
is real symmetric and depends on a vector of two parameters
$\mathbf{p} = (s_1,s_2)$ (see \cite{Berry2} for other ways of
introducing two parameters). The third component of the direction
vector $\mathbf{s}$ is found as $s_3 = \pm\sqrt{1-s_1^2-s_2^2}$,
where the cases of two different signs should be considered
separately. Below we assume that three dielectric constants
$\eta_1
> \eta_2 > \eta_3$ are different. This corresponds to biaxial anisotropic
crystals.

The nonzero eigenvalues $\lambda_\pm$ of the matrix
$\mathbf{A}(\mathbf{p})$ are found explicitly in the form
\cite{Lewin}
    \begin{equation}
    \label{eqE.5b}
    \lambda_\pm = \frac{\mathrm{trace}\,\mathbf{A}}{2}
    \pm\frac{1}{2}\sqrt{2\,\mathrm{trace}\,(\mathbf{A}^2)
    -(\mathrm{trace}\,\mathbf{A})^2}.
    \end{equation}
The eigenvalues $\lambda_\pm$ are the same for opposite directions
$\mathbf{s}$ and $-\mathbf{s}$. By using (\ref{eqE.2}) and
(\ref{eqE.5}) in (\ref{eqE.5b}), it is straightforward to show
that two eigenvalues $\lambda_+$ and $\lambda_-$ couple at
    \begin{equation}
    \label{eqE.6}
    \mathbf{s}_0
    = (S_1,S_2,S_3),\
    \lambda_0 = \eta_2;\ \
    S_1 = \pm\sqrt{(\eta_1-\eta_2)/(\eta_1-\eta_3)},\
    S_2 =0,\
    S_3 = \pm\sqrt{1-S_1^2},
    \end{equation}
which determine four diabolic points (for two signs of $S_1$ and
$S_3$), also called optic axes \cite{RR}. The double eigenvalue
$\lambda_0 = \eta_2$ of the matrix $\mathbf{A}_0 =
\mathbf{A}(\mathbf{p}_0)$, $\mathbf{p}_0 = (S_1,0)$ possesses two
eigenvectors
    \begin{equation}
    \label{eqE.7}
    \mathbf{u}_1 = \left(\begin{array}{c}
        0 \\ 1 \\ 0
    \end{array}\right),\quad
    \mathbf{u}_2 = \left(\begin{array}{c}
        S_3 \\ 0 \\ -S_1
    \end{array}\right),
    \end{equation}
satisfying normalization conditions (\ref{eq0.1b}). Using
expressions (\ref{eqE.5}) and (\ref{eqE.7}), we evaluate the
vectors $\mathbf{f}_{ij}$ with components (\ref{eq0.4}) for optic
axes (\ref{eqE.6}) as
    \begin{equation}
    \label{eqE.8}
    \mathbf{f}_{11} = (0,\,0),\quad
    \mathbf{f}_{22} = (2(\eta_3-\eta_1)S_1,\,0),\quad
    \mathbf{f}_{12} = \mathbf{f}_{21}
    = (0,\,(\eta_3-\eta_1)S_1S_3).
    \end{equation}
By using (\ref{eqE.6}) and (\ref{eqE.8}) in (\ref{eq1.1}), we
obtain the local asymptotic expression for the cone singularities
in the space $(s_1,s_2,\lambda)$ as
    \begin{equation}
    \label{eqE.9}
    (\lambda-\eta_2-(\eta_3-\eta_1)S_1(s_1-S_1))^2
    = (\eta_3-\eta_1)^2S_1^2((s_1-S_1)^2
    +S_3^2s_2^2).
    \end{equation}
Equation (\ref{eqE.9}) is valid for each of the four optic axes
(\ref{eqE.6}).

\begin{figure}
    \begin{center}
    \includegraphics[angle=0, width=0.55\textwidth]{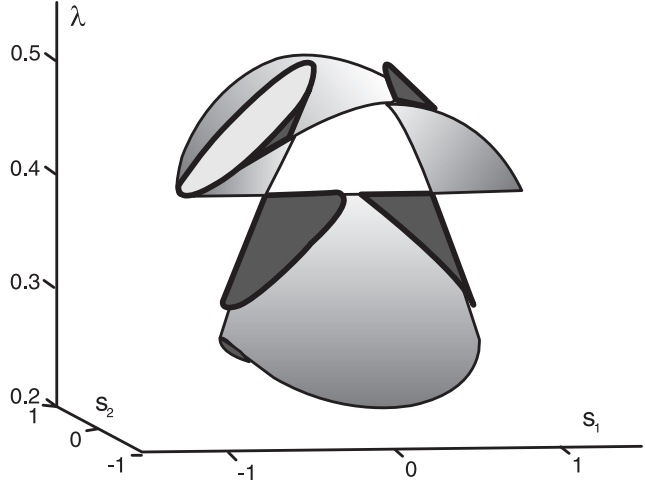}
    \end{center}
    \caption{Diabolic singularities near optic axes and
    their local approximations.}
    \label{figE.1}
    \end{figure}

As an example, consider the case of $\eta_1 = 0.5$, $\eta_2 =
0.4$, $\eta_3 = 0.1$. Conical surfaces (\ref{eqE.9}) are shown in
Figure~\ref{figE.1} together with the exact eigenvalue surfaces
(\ref{eqE.5b}). The two optic axes presented in
Figure~\ref{figE.1} are $\mathbf{s}_0 = (\pm1/2,\,0,\,\sqrt{3}/2)$
with the double eigenvalue $\lambda_0=2/5$; the eigenvalue
surfaces for the opposite directions $\mathbf{s}_0 =
(\pm1/2,\,0,\,-\sqrt{3}/2)$ are exactly the same.


Now let us assume that the crystal possesses absorption and
chirality. Then the matrix family (\ref{eqE.5}) takes a complex
perturbation
$\mathbf{A}(\mathbf{p})+\Delta\mathbf{A}(\mathbf{p})$, where
    \begin{equation}
    \label{eqE.10}
    \Delta\mathbf{A}(\mathbf{p}) =
    (\mathbf{I}-\mathbf{s}\mathbf{s}^T)
    ({\boldsymbol\eta}_{dichroic}+{\boldsymbol\eta}_{chiral})
    (\mathbf{I}-\mathbf{s}\mathbf{s}^T).
    \end{equation}
Assume that the absorption and chirality are weak, i.e.,
$\varepsilon =
\|{\boldsymbol\eta}_{dichroic}\|+\|{\boldsymbol\eta}_{chiral}\|$
is small. Then we can use asymptotic formulae of Sections 2 and 3
to describe unfolding of diabolic singularities of the eigenvalue
surfaces. For this purpose, we need to know only the value of the
perturbation $\Delta\mathbf{A}$ at the optic axes of the
transparent non-chiral crystal $\mathbf{s}_0$.

Substituting matrix (\ref{eqE.10}) evaluated at optic axes
(\ref{eqE.6}) into expression (\ref{eq0.8}), we obtain
    \begin{equation}
    \label{eqE.11}
    \begin{array}{c}
        \varepsilon_{11} = i\eta^d_{22},\quad
        \varepsilon_{22} =
        i\eta^d_{11}S_3^2-2i\eta^d_{13}S_1S_3
        +i\eta^d_{33}S_1^2,\\[7pt]
        \varepsilon_{12} = -i(\eta^d_{23}
        +\gamma_{11}S_1+\gamma_{13}S_3)S_1+
        i(\eta^d_{12}-\gamma_{13}S_1-\gamma_{33}S_3)S_3,\\[7pt]
        \varepsilon_{21} = -i(\eta^d_{23}
        -\gamma_{11}S_1-\gamma_{13}S_3)S_1+
        i(\eta^d_{12}+\gamma_{13}S_1+\gamma_{33}S_3)S_3.
    \end{array}
    \end{equation}
By using formulae (\ref{eq1.5}), we get
    \begin{equation}
    \label{eqE.12}
    \begin{array}{l}
        \mu = i(\eta^d_{22}+\eta^d_{11}S_3^2-2\eta^d_{13}S_1S_3
        +\eta^d_{33}S_1^2)/2,\\[7pt]
        \xi = i(\eta^d_{22}-\eta^d_{11}S_3^2+2\eta^d_{13}S_1S_3
        -\eta^d_{33}S_1^2)/2,\\[7pt]
        \eta = i(\eta^d_{12}S_3-\eta^d_{23}S_1),\\[7pt]
        \zeta = -i\left(\gamma_{11}S_1^2+2\gamma_{13}S_1S_3
        +\gamma_{33}S_3^2\right).
    \end{array}
    \end{equation}
We see that $\mu$, $\xi$, and $\eta$ are purely imaginary numbers
depending only on dichroic properties of the crystal (absorption).
The quantity $\zeta$ depends only on chiral properties of the
crystal; $\zeta$ is purely imaginary if the optical activity
tensor $\boldsymbol\gamma$ is real.

Singularities for crystals with weak dichroism and chirality were
studied recently in~\cite{Berry2}. It was shown that the double
coffee filter singularity arises in absorption-dominated crystals,
and the sheets of real parts of eigenvalues are separated in
chirality-dominated crystals. According to the results of
Section~3, these two cases are explicitly determined by the
conditions $D > 0$ and $D < 0$, respectively, where $D =
\mathrm{Im}^2\xi+\mathrm{Im}^2\eta-\mathrm{Im}^2\zeta$. These
conditions are new and important because they provide quantitative
definitions of absorption-dominated and chirality-dominated
regimes for unfolding of the diabolic singularity in terms of
components of the inverse dielectric tensor. Indeed, according to
(\ref{eqE.12}), $\xi$ and $\eta$ depend linearly on all the
components of the tensor ${\boldsymbol\eta}_{dichroic}$, while
$\zeta$ depends linearly on the components $\gamma_{ij}$, $i,j =
1,3$ of the optical activity tensor $\boldsymbol\gamma$.

Note that according to the sign of the quantity $D$ taken at
different optic axes we can classify crystals by their optic
properties. For example, the important case is a
chirality-dominated crystal with $D<0$ for all four optic axes.
Then real parts of the eigenvalues separate for all directions
${\bf s}$.

There are four optic axes (\ref{eqE.6}), which determine two pairs
of opposite space direction $\pm\mathbf{s}_0$. It is easy to see
that the unfolding conditions coincide for the optic axes given by
opposite directions, while these conditions are different for
different pairs of optic axes. In the absorption-dominated case,
when diabolic singularities unfold into coffee-filters near two
opposite optic axes $\pm\mathbf{s}_0 = \pm(S_1,0,S_3)$, the four
exceptional points of eigenvalue coupling $\pm\mathbf{s}_a$ and
$\pm\mathbf{s}_b$ (also called singular axes) appear. By using
(\ref{eqE.8}) in (\ref{eq1.4}), we obtain the asymptotic formulae
    \begin{equation}
    \label{eqE.13}
    s_1^{a,b} = S_1+\frac{x_{a,b}}{(\eta_1-\eta_3)S_1},\quad
    s_2^{a,b} = \frac{y_{a,b}}{(\eta_3-\eta_1)S_1S_3},\quad
    s_3^{a,b} = \sqrt{1-{(s_1^{a,b})}^2-{(s_2^{a,b})}^2},
    \end{equation}
for the components of the vectors $\mathbf{s}_{a,b}$, where
$x_{a,b}$ and $y_{a,b}$ are found by using expressions
(\ref{eq1.14}), (\ref{eq1.15}), and (\ref{eqE.12}). In particular,
for non-chiral crystals, we have $\zeta = 0$. Then expressions
(\ref{eq1.14}), (\ref{eq1.15}) yield $x_{a,b} =
\pm\mathrm{Im}\,\eta = \eta^d_{12}S_3-\eta^d_{23}S_1$ and $y_{a,b}
= \mp\mathrm{Im}\,\xi =
(\eta^d_{22}-\eta^d_{11}S_3^2+2\eta^d_{13}S_1S_3
-\eta^d_{33}S_1^2)/2$.

Equation $\mathrm{Im}\,c = 0$ determines a line of singularities
in the parameter space $\mathbf{p} = (s_1,s_2)$. By using
(\ref{eqE.8}), (\ref{eqE.12}) in (\ref{eq1.4}), (\ref{eq1.8}), we
find this line in the form
    \begin{equation}
    \label{eqE.14}
    (s_1-S_1)S_1(\eta_1-\eta_3)\mathrm{Im}\,\xi
    -s_2S_1S_3(\eta_1-\eta_3)\,\mathrm{Im}\,\eta
    -\mathrm{Re}\,\zeta\,\mathrm{Im}\,\zeta = 0.
    \end{equation}
In the absorption-dominated case, line (\ref{eqE.14}) contains two
exceptional points $\mathbf{p}_{a,b} = (s_1^{a,b},s_2^{a,b})$
corresponding to the singular axes $\mathbf{s}_{a,b}$. A segment
between the points $\mathbf{p}_a$ and $\mathbf{p}_b$ corresponds
to the coincidence of real parts of the eigenvalues
$\mathrm{Re}\,\lambda_+ = \mathrm{Re}\,\lambda_-$, while imaginary
parts of the eigenvalues $\mathrm{Im}\,\lambda_+ =
\mathrm{Im}\,\lambda_-$ merge at points of line (\ref{eqE.14})
outside this segment, see Figure~\ref{fig2}b. In the
chirality-dominated case, when singular axes do not appear,
imaginary parts of the eigenvalues $\mathrm{Im}\,\lambda_+ =
\mathrm{Im}\,\lambda_-$ coincide at points of the whole line
(\ref{eqE.14}), see Figure~\ref{fig2}a. If the optical activity
tensor $\boldsymbol\gamma$ is real or purely imaginary, then the
line of singularities (\ref{eqE.14}) passes through the diabolic
point $\mathbf{p}_0$, and position of this line does not depend on
$\boldsymbol\gamma$.


    \begin{figure}
    \begin{center}
    \includegraphics[angle=0, width=0.74\textwidth]{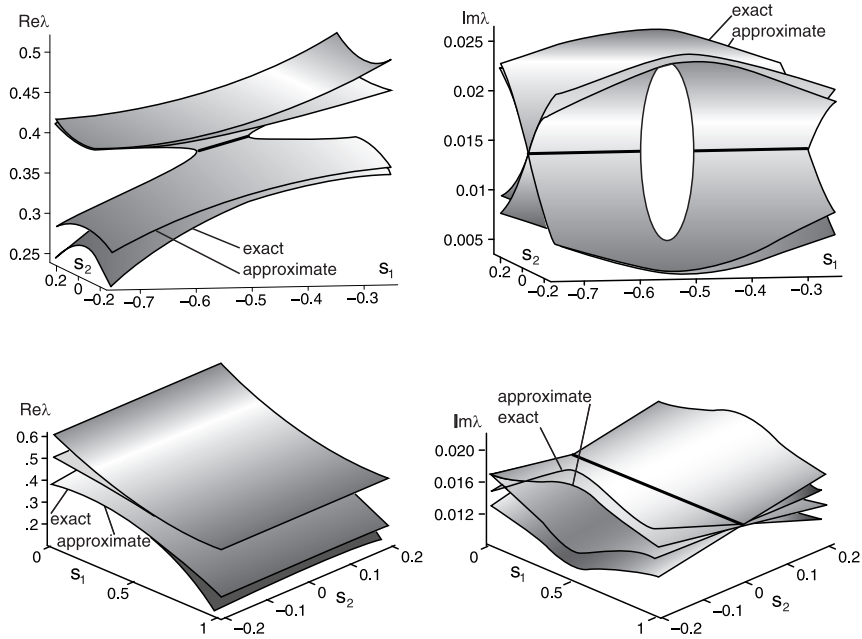}
    \end{center}
    \caption{Unfolding of singularities near optic
    axes.}
    \label{figE.2}
    \end{figure}

As a numerical example, let us consider a crystal possessing weak
absorption and chirality described by the tensors (\ref{eqE.1b}),
(\ref{eqE.1c}) with
    \begin{equation}
    \label{eqE.15}
    {\boldsymbol\eta}_{dichroic}
    = \frac{i}{200}\left(\begin{array}{ccc}
         3 & 2 & 0 \\
         2 & 3 & 1 \\
         0 & 1 & 3
    \end{array}\right),\quad
    {\boldsymbol\gamma}
    = \frac{1}{200}\left(\begin{array}{ccc}
         3 & 1 & 2 \\
         1 & 3 & 1 \\
         2 & 1 & 3
    \end{array}\right).
    \end{equation}
A corresponding transparent non-chiral crystal is characterized by
$\eta_1 = 0.5$, $\eta_2 = 0.4$, $\eta_3 = 0.1$, and its eigenvalue
surfaces with two optic axes are presented in Figure~\ref{figE.1}.
By using (\ref{eqE.15}) in (\ref{eqE.12}), we find that the
condition $D = \frac{7}{160000}(4\sqrt{3}-5) > 0$ is satisfied for
the left optic axis $\mathbf{s}_0 = (-1/2,\,0,\,\sqrt{3}/2)$.
Hence, the diabolic singularity bifurcates into a double coffee
filter with two exceptional points whose coordinates according to
expressions (\ref{eqE.13}) are
    \begin{equation}
    {\bf p}_a=\left(
    -\frac{1}{2}-\frac{1}{80}\sqrt{-35+28\sqrt{3}}, 0
    \right),~~
    {\bf p}_b=\left(
    -\frac{1}{2}+\frac{1}{80}\sqrt{-35+28\sqrt{3}}, 0
    \right).
    \label{eqE.16a}
    \end{equation}
Local approximations of the eigenvalue surfaces are given by
expressions (\ref{eq1.9}), (\ref{eq1.10}), where
    \begin{equation}
    \label{eqE.16}
    \mathrm{Re}\,c =
    \frac{35-28\sqrt{3}}{160000}+\frac{1}{25}(s_1+1/2)^2+\frac{3}{100}s_2^2,\ \
    \mathrm{Im}\,c = -\frac{6+\sqrt{3}}{2000}\,s_2.
    \end{equation}
Figure~\ref{figE.2}a shows these local approximations compared
with the exact eigenvalue surfaces given by (\ref{eqE.5b}). For
the right optic axis $\mathbf{s}_0 = (1/2,\,0,\,\sqrt{3}/2)$, the
condition $D = -\frac{7}{160000}(4\sqrt{3}+5) < 0$ is satisfied.
Hence, the eigenvalue sheets (for real parts) separate under the
bifurcation of the right diabolic singularity. Approximate and
exact eigenvalue surfaces are shown in Figure~\ref{figE.2}b. The
approximations are given by expressions (\ref{eq1.9}),
(\ref{eq1.10}), where
    \begin{equation}
    \label{eqE.17}
    \mathrm{Re}\,c =
    \frac{35+28\sqrt{3}}{160000}+\frac{1}{25}(s_1-1/2)^2+\frac{3}{100}s_2^2,\ \
    \mathrm{Im}\,c = -\frac{6-\sqrt{3}}{2000}\,s_2.
    \end{equation}
We observe that the unfolding types are different for different
optic axes. As it is seen from Figure~\ref{figE.2}, the asymptotic
formulae provide an accurate description for unfolding of
eigenvalue surfaces near diabolic points.

\section{Conclusion}

Non-Hermitian Hamiltonians and matrices usually appear in physics
when dissipative and other non-conservative effects are taken into
account. The known examples are complex refractive indices in
optics and complex potentials describing the scattering of
electrons or X-rays. Traditionally, non-Hermitian matrices appear
in physics as a perturbation of Hermitian matrices. As it is
stated in \cite{Berry}, Hermitian physics differs radically from
non-Hermitian physics in case of coalescence (coupling) of
eigenvalues. In the present paper we have studied this important
case carefully. We gave analytical description for unfolding of
eigenvalue surfaces due to an arbitrary complex perturbation with
the singularities known in the literature as a "double coffee-
filter" and a "diabolic circle". We emphasize that the developed
theory requires only eigenvectors and derivatives of the matrices
taken at the singular point, while the size of the matrix and its
dependence on parameters are arbitrary. This makes the presented
theory powerful and practical for a wide class of physical
problems. The given physical example from crystal optics
demonstrates applicability and accuracy of the theory.

\section{Acknowledgement}

The work is supported by the research grants RFBR-NSFC
02-01-39004, RFBR 03-01-00161, and CRDF-BRHE Y1-MP-06-19.

\end{document}